\documentclass[aps,prl,twocolumn,superscriptaddress]{revtex4-2}

\usepackage{amsmath,amssymb,graphicx}

\begin{document}

\title{Universal inverse-cube thickness scaling of projectile penetration energy in ultrathin films}

\author{Alessio Zaccone}
\affiliation{Department of Physics ``A. Pontremoli", University of Milan, via Celoria 16, 20133 Milan, Italy}

\author{Timothy W. Sirk}
\affiliation{Polymers Branch, US Army DEVCOM Army Research Laboratory, Aberdeen Proving Ground, Maryland 21005, USA}

\begin{abstract}
Ultrathin films of widely different materials exhibit a dramatic enhancement
of projectile penetration resistance under high--velocity impact.
Despite extensive simulations and experiments, a unifying physical explanation
has remained elusive.
Here we show that the thickness dependence of the specific penetration energy
obeys a universal law,
$E_p^*(h)=E_{p,\infty}^*+B h^{-3}$,
independent of chemical composition and degree of disorder.
The inverse--cube scaling is traced back to a finite--size correction to the
effective shear modulus arising from the suppression of long--wavelength
nonaffine deformation modes in confined solids.
The scaling quantitatively describes impact data for multilayer graphene, graphene oxide, and polymer thin films, revealing a common elastic origin for nanoscale impact resistance.
\end{abstract}

\maketitle

Under high--velocity impact in the elastic--inertial regime, projectile
resistance is governed by stress-wave propagation and momentum transfer
within the target.
The relevant material parameter controlling the attainable transient shear
stresses is the effective high-rate shear modulus.
For nanometric films, this modulus is strongly affected by confinement, which
suppresses long--wavelength deformation modes.
In real (disordered and partially ordered) solids, these modes are predominantly
nonaffine and contribute negatively to the shear modulus.

These modes correspond to internal atomic or molecular motions and do not involve global plate or beam bending. In classical plate theory the bending stiffness scales as $D\propto Eh^3$, implying that thinner films bend more easily and would therefore predict the opposite thickness trend. The $h^{-3}$ scaling proposed here instead arises from confinement-induced suppression of long-wavelength nonaffine shear modes \cite{Kostya}.

The mechanical response of real-world solids differs fundamentally from that
of perfect crystalline materials, due to the presence of nonaffine particle
displacements.
At the microscopic level, the equation of motion for the displacement $\mathbf{x}$ of particle $i$ under an
applied strain can be derived from a system--bath Caldeira--Leggett Hamiltonian
and reads \cite{MilkusZaccone2017,Zaccone_book}
\begin{equation}
m \ddot{\mathbf{x}}_i + \nu \dot{\mathbf{x}}_i + \sum_j H_{ij} \mathbf{x}_j = \mathbf{\Xi}_{i,k\ell}\,\eta_{k\ell},
\label{eq:nonaffine_eom}
\end{equation}
where $m$ is the particle mass, $\nu$ an interatomic friction coefficient,
$H_{ij}$ the Hessian (dynamical matrix) describing harmonic nearest-neighbor
interactions, $\eta_{k\ell}$ the applied strain tensor ($k\ell = xy$ for shear), and
$\mathbf{\Xi}_{i,k\ell}$ the nonaffine force field generated by the lack of local
inversion symmetry (due to defects, thermal motion or disorder).
In real systems, the force on each atom does not vanish in an affine deformation, requiring additional nonaffine relaxations that reduce the shear modulus \cite{Scossa}.

Solving Eq.~(\ref{eq:nonaffine_eom}) in Fourier space and projecting onto the eigenvectors of the Hessian matrix, further manipulations yield the complex shear
modulus \cite{MilkusZaccone2017,lemaitre2006}
\begin{equation}
G^*(\omega) = G_\infty - A
\sum_{\mathbf{k},\lambda}
\frac{\omega_{\mathbf{k},\lambda}^2}
{\omega_{\mathbf{k},\lambda}^2 - \omega^2 + i \omega \nu},
\label{eq:complex_modulus}
\end{equation}
where $G_\infty$ is the affine (Born) shear modulus that survives in the infinite frequency limit, $\omega_{\mathbf{k},\lambda}$
are the eigenfrequencies of the vibrational modes indexed by wavevector
$\mathbf{k}$ and polarization $\lambda$, and the second term represents the
softening contribution from nonaffine relaxations.
Following nonaffine lattice dynamics \cite{MilkusZaccone2017,Zaccone_book,lemaitre2006},
the complex shear modulus can be written as a sum over vibrational eigenmodes
\begin{equation}
G^*(\omega)=G_\infty
-\frac{1}{V}\sum_{p}\frac{\Gamma_p(\omega_p)}{\omega_p^{\,2}-\omega^2+i\omega\nu},
\label{eq:Gstar_modes}
\end{equation}
where $G_\infty$ is the affine (Born) modulus, $\omega_p$ are eigenfrequencies,
$\nu$ is a microscopic damping coefficient, and $\Gamma_p\ge 0$ are mode
coupling weights (set by projections of the nonaffine force field onto mode
$p$).
In the quasistatic limit $\omega\to 0$ one obtains
$G'(\omega\!\to\!0)=G_\infty
-G_{NA}$ where the extent of the residual nonaffine contribution $G_{NA}$ depends on the type of chemical bonding and/or molecular forces, and on the atomic-scale geometric environment (it tends to vanish only for perfectly centrosymmetric crystals at low temperature).

We consider a thin film of thickness $L\equiv L_z$ with lateral dimensions
$L_x,L_y\gg L$ (i.e. $L\equiv L_z \ll L_x,L_y$).
In a finite system the allowed wavevectors are discrete, and the smallest
accessible magnitude is set by the system dimensions. In particular,
\begin{equation}
k_{\min}\equiv |{\bf k}_{\min}|
=2\pi\sqrt{\left(\frac{1}{L_x}\right)^2+\left(\frac{1}{L_y}\right)^2+\left(\frac{1}{L}\right)^2}
\;\sim\;\frac{1}{L},
\label{eq:kmin}
\end{equation}
so that in the thin-film limit the infrared cut-off is controlled by the
thickness. For a rigorous justification of this relation, including the exact prefactors, see \cite{Phillips2021}.

Following Refs.~\cite{Kostya, Phillips2021}, we replace the discrete
sum over modes by a momentum integral. At low frequencies the nonaffine
softening contribution can be represented in the form
\begin{equation}
G' = G_\infty - \alpha \int_{k_{\min}}^{k_D} k^2\,dk,
\label{eq:Gprime_integral}
\end{equation}
where $G_\infty$ is the affine (Born) modulus, $k_D$ is a Debye-like ultraviolet
cutoff, and $\alpha>0$ collects material-dependent prefactors (including
mode-coupling weights). Performing the integral gives
\begin{equation}
G' = G_\infty - \frac{\alpha}{3}\left(k_D^3-k_{\min}^3\right)
= \underbrace{\left(G_\infty-\frac{\alpha}{3}k_D^3\right)}_{\equiv\,G'_{\rm bulk}}
+\frac{\alpha}{3}k_{\min}^3.
\label{eq:Gprime_eval}
\end{equation}
Using Eq.~(\ref{eq:kmin}), $k_{\min}\sim 1/L$, one obtains the inverse-cube
finite-size correction
\begin{equation}
G'(L)=G'_{\rm bulk}+\frac{\beta}{3}\,L^{-3},
\qquad \beta\equiv\alpha,
\label{eq:Gprime_Lm3}
\end{equation}
which is the explicit confinement-induced $L^{-3}$ scaling reported in
Eq.~(6) of Phillips \textit{et al.}~\cite{Phillips2021}.
Identifying $L$ with the film thickness $h$ yields
\begin{equation}
G'(h)=G'_{\rm bulk}+\beta'\,h^{-3},
\label{eq:Gprime_hm3}
\end{equation}
with $\tilde{\beta}=\beta/3$ after absorbing numerical factors into the
material-dependent prefactor.

In the elastic--inertial regime of high-velocity penetration, projectile
resistance is governed by stress-wave propagation and momentum transfer
within the target. This framework does not attempt to replace ballistic-limit
models that include strength and failure criteria (e.g., Phoenix-type models),
but instead identifies the physical origin of the observed thickness scaling
within the elastic stiffness entering such models.
The characteristic stress scale generated around the projectile boundary
scales with the transverse wave impedance.
The transverse sound speed satisfies
$c_T \sim \sqrt{G'/\rho}$,
and the corresponding impedance is
$Z \sim \rho c_T \sim \sqrt{\rho G'}$.
For comparable projectile geometry and impact velocity,
the attainable shear stress and the rate of momentum transfer therefore
inherit the thickness dependence of $G'(h)$.
Consequently, the thickness dependence of the specific penetration energy
follows that of the shear modulus,
so that $E_p^*(h)\propto G'(h)$
\cite{Zukas1982,GoldsmithImpact,Florence1967}.
Combining with Eq.~(\ref{eq:Gprime_hm3}) gives
\begin{equation}
E_p^*(h)=E_{p,\infty}^*+B\,h^{-3},
\label{eq:Ep_hm3}
\end{equation}
where $E_{p,\infty}^*\propto G'_{\rm bulk}$ and $B\propto \tilde{\beta}/\rho$.
In practice, microscopic length scales introduce a small cutoff $h_0$, leading to a regularized form $E_p^*(h)=E_{p,\infty}^*+B(h+h_0)^{-3}$ that avoids divergence as $h\rightarrow0$ \cite{Pugno_2007}.

We emphasize that $E_p^*$ is not a velocity-independent intrinsic material constant but an experimentally reported impact metric that depends on test conditions and velocity regime. In high-velocity impact of thin films, projectile arrest typically occurs via localized plugging and rapid momentum transfer around the projectile boundary rather than global bending. The present theory therefore links the observed thickness scaling of $E_p^*$ to confinement-induced modification of the high-rate shear stiffness.
Equation~(\ref{eq:Ep_hm3}) provides a direct link
between nonaffine elasticity theory and the experimentally observed enhancement
of penetration resistance in ultrathin films.

\begin{figure*}[t]
\centering
\includegraphics[width=0.95\textwidth]{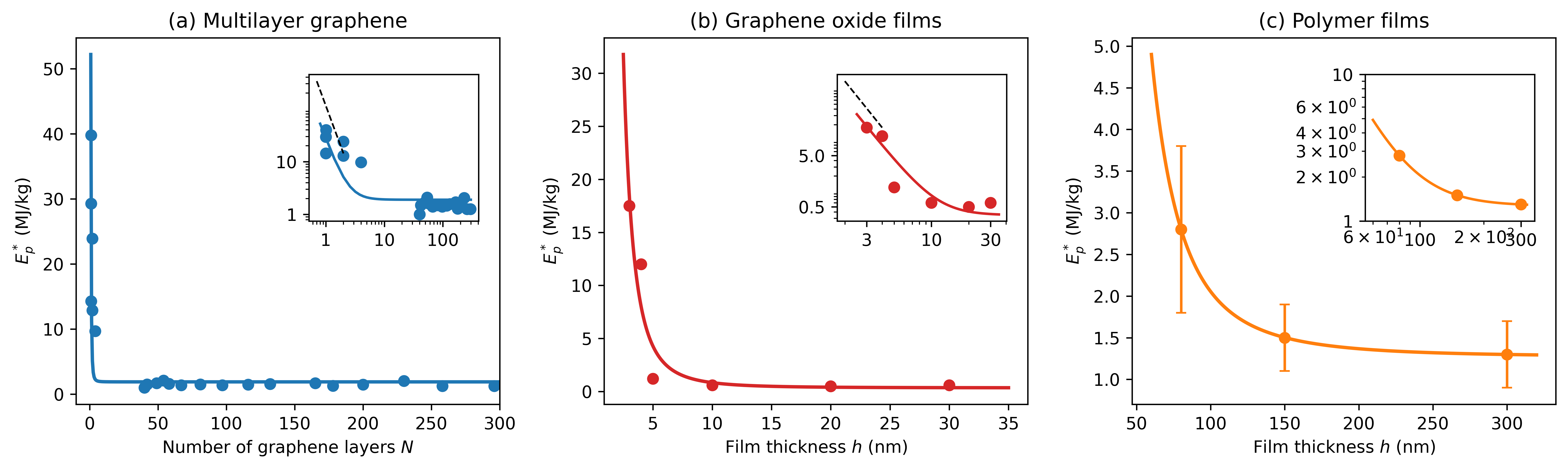}
\caption{
Universal inverse--cube thickness scaling of the specific penetration energy.
(a)~Multilayer graphene: specific penetration energy $E_p^*$ as a function of
the number of graphene layers $N$, digitized from Fig.~4b of Bizao
\textit{et al.}~\cite{Bizao2018}.
(b)~Graphene oxide films under $1000~\mathrm{m/s}$ impact~\cite{ImpactData2025}.
(c)~Polymer thin films under $800~\mathrm{m/s}$ impact from \cite{Hyon2018}.
Solid lines are fits to Eq.~\eqref{eq:Ep_hm3}. Dashed lines in the log-log insets indicate the pure power-law trend $h^{-3}.$
}
\label{fig:universal_scaling}
\end{figure*}

We test Eq.~(\ref{eq:Ep_hm3}) against ballistic data for multilayer graphene \cite{Bizao2018,Lee2014Science,Yoon2016Carbon,Haque2016Carbon}, graphene oxide films under $1000~\mathrm{m/s}$ impact \cite{ImpactData2025}, and polymer thin films at comparable velocities \cite{Hyon2018} (Fig.~\ref{fig:universal_scaling}). In all cases the model captures both the strong enhancement of penetration resistance in the ultrathin regime and the crossover toward a bulk plateau, supporting a universal finite-size elastic mechanism.

The coefficient $B$ measures the strength of confinement--induced stiffening.
Its magnitude determines the thickness at which the finite--size contribution
$B h^{-3}$ overtakes the bulk term $E_{p,\infty}^*$.
For graphene oxide films, the fitted value implies that already at
$h\sim5~\mathrm{nm}$ the finite--size contribution exceeds the bulk penetration
energy by more than an order of magnitude.

\begin{table}[t]
\centering
\caption{Fitted values of the finite-size coefficient $B$ in
$E_p^*(x)=E_{p,\infty}^*+B x^{-3}$ for the three systems shown in Fig.~\ref{fig:universal_scaling}.}
\begin{tabular}{lccc}
\hline
System & Variable $x$ & $B$ & Units \\
\hline
Multilayer graphene & $N$ &
$2.6\times10^{1}$ & MJ/kg \\
Graphene oxide films & $h$ &
$4.9\times10^{2}$ & MJ\,nm$^{3}$/kg \\
Polymer thin films & $h$ &
$7.8\times10^{5}$ & MJ\,nm$^{3}$/kg \\
\hline
\end{tabular}
\end{table}

This behavior is predicted by microscopic elasticity: long-wavelength nonaffine modes soften the shear modulus in disordered solids \cite{Zaccone_book}, hence their suppression due to confinement increases the rigidity. Experiments on ultrathin polymer films likewise report strong thickness-dependent elastic moduli \cite{Stafford2004NatMat,Stafford2006Macromolecules}, consistent with confinement restricting accessible deformation modes. Suppressing these long-wavelength modes produces the $h^{-3}$ stiffening highlighted here.

The scaling $E_p^*(h)=E_{p,\infty}^*+B h^{-3}$ is expected to hold most clearly in the elastic--inertial high-rate regime. At sufficiently high impact velocities, the loading time becomes shorter than structural relaxation times associated with molecular rearrangements, interlayer sliding, or chain mobility. The response therefore becomes effectively solid-like, and penetration resistance is controlled primarily by shear stiffness and acoustic impedance, consistent with $E_p^*(h)\propto G'(h)$. In this regime projectile arrest occurs through localized plugging and rapid momentum transfer mediated by transient shear stresses rather than global bending. At lower velocities, viscoelastic or plastic relaxation mechanisms may mask the purely geometric confinement effect and produce deviations from simple power-law scaling.

In conclusion, the inverse--cube thickness dependence of projectile penetration energy (followed by the bulk-like plateau) is shown
to be a universal consequence of finite-size elastic stiffening due to confinement-induced suppression of
long-wavelength nonaffine modes.
By linking ballistic impact resistance to microscopic nonaffine elasticity, the present
results unify disparate experimental observations across crystalline,
amorphous, and polymeric thin films within a single, quantitative framework.

\section*{Acknowledgments}
AZ acknowledges funding from the European Union through Horizon Europe ERC Grant number: 101043968 “Multimech” and from US Army Research Office through contract nr.  W911NF-22-2-0256.

\bibliography{refs}

\end{document}